\documentclass{optica-article}

\journal{opticajournal} 

\articletype{Research Article}

\usepackage{subcaption}
\usepackage{graphicx}

\begin{document}

\title{3D in-situ profiling in a laser micromachining station using dual-comb LiDAR}

\author{Hayk Soghomonyan,\authormark{1,*} Justinas Pupeikis,\authormark{1,2}, Benjamin Willenberg\authormark{1,2}, Armin Stumpp,\authormark{3}, Lukas Lang\authormark{1,2}, Christopher R. Phillips,\authormark{1}, Bojan Resan\authormark{3} and Ursula Keller\authormark{1}}

\address{\authormark{1}Department of Physics, Institute for Quantum Electronics, ETH Zurich, 8093 Zurich, Switzerland\\
\authormark{2}K2 Photonics AG, Hardturmstrasse 161, 8005 Zurich, Switzerland\\
\authormark{3}School of Engineering, University of Applied Sciences and Arts Northwestern Switzerland, Klosterzelgstrasse 2, 5210 Windisch, Switzerland}

\email{\authormark{*}hayks@phys.ethz.ch} 


\begin{abstract*} 
One of the main challenges in laser micromachining is the complexity of the process development. The available monitoring capabilities are mostly limited to external measurement systems, significantly slowing down this step. We address this limitation by integrating a coaxial dual-comb LiDAR system directly into a laser micromachining station, enabling in-situ, non-destructive 3D profiling without removing the workpiece. The sub-micron raw axial precision of the system, the operation without mechanical delay scanning and a large working range make the system well-suited for profiling a large variety of micromachined structures. By providing in-situ feedback this approach will significantly accelerate laser micromachining process development.

\end{abstract*}

\section{Introduction}
Laser micromachining has emerged as the key technique for producing high-precision components for industries such as consumer electronics, medical, watch and jewelry, automotive, aerospace, and defense\cite{chichkov_femtosecond_1996, neuenschwander_processing_2010, schoonderbeek_laser_2010}. However, the development and qualification of suitable micromachining processes remain time-intensive and require significant expertise \cite{neuenschwander_laser_2016,moglia_new_2021, yildirim_development_2024}. Development typically involves multiple iterations of micromachining a part, followed by inspection with an external device, such as a laser scanning microscope or a white light interferometer. This workflow necessitates moving and remounting parts for inspection, which degrades precision, repeatability, and prolongs qualification times. Moreover, even established processes can fail due to variations in material properties or machining laser performance, necessitating requalification.
The integration of in-situ non-destructive testing (NDT) and evaluation is crucial for establishing an active feedback loop that ensures the achievement of the desired 3D profile, irrespective of variations in sample or machining laser quality. In addition, in-situ profiling could open opportunities for machine learning deployment in the micromachining process, which could translate into significant time savings for the end users. One approach is to incorporate an optical coherence tomography (OCT) system into the micromachining station, as demonstrated in previous reports \cite{huang_optical_1991,massow_oct-aided_2009,wiesner_optical_2010, stehmar_inline_2022}. Both spectral-domain OCT (SD-OCT) and swept-source OCT (SS-OCT) offer excellent performance that has led to their widespread deployment, especially in ophthalmology \cite{varghese_revolutionizing_2024}.  However, high-resolution OCT systems are expensive and would require significant customization to meet precision (sub-micrometer), measurement depth (millimeter to centimeter scale), wavelength (close to the machining laser) and integration constraints applicable for deployment in laser materials processing workstations \cite{huang_optical_1991,massow_oct-aided_2009,wiesner_optical_2010, stehmar_inline_2022}.

White-light interferometry (WLI) is also a suitable technique. However, the need for mechanical delay scanning may not be compatible with the rapid beam scanners typically employed in micromachining. Additional significant integration challenges emerge because the reference path needs to be introduced, and optical delay over large distances, typically tens to hundreds of centimeters, needs to match, which is sensitive to environmental drifts. The frequency comb-based WLI \cite{maslowski_surpassing_2016} overcomes the range limitations of incoherent light sources, but still involves mechanical scanning.

Here, we demonstrate an alternative approach utilizing the emerging dual-comb ranging technique \cite{lee_ultrahigh_2001, coddington_rapid_2009}, which has compelling advantages such as high precision, long working range, and rapid optical delay scanning, especially when utilizing gigahertz dual optical frequency combs \cite{lee_ultrahigh_2001,link_dual-comb_2015,mohler_dual-comb_2017,voumard_1-ghz_2022, phillips_coherently_2023, camenzind_long-range_2025}. The use of a shared laser cavity arrangement also helps reduce noise \cite{link_dual-comb_2015,pupeikis_spatially_2022} and enables a compact and cost-effective footprint \cite{link_dual-comb_2015,willenberg_thz-tds_2024}.

The working principle of dual-comb ranging relies on two optical frequency combs with slightly different repetition rates. The pulse train reflected from the measurement surface is interfered with a second comb acting as the local oscillator. As the two pulse trains gradually walk through each other, the interference signal (interferogram) appears in the time domain and can be directly measured with available electronics. This time-domain sampling yields interferograms that preserve time-of-flight (ToF) information with high precision, effectively slowed down by the factor $\frac{f_{rep}}{\Delta f_{rep}}$, while eliminating the need for mechanical delay scanning. Here, $f_{rep}$ is the repetition rate of the laser and $\Delta f_{rep}$ the small difference in the repetition rates between the two combs. The long mutual coherence of the pulse trains enables dual-comb LiDAR (light detection and ranging) to achieve measurement ranges of kilometers, while $\Delta f_{rep}$ sets the effective acquisition rate of the ranging measurement and can be tuned from 0 Hz up to the hundreds of kHz scale depending on the system configuration.

In contrast to commonly used SD-OCT \cite{wojtkowski_vivo_2002}, dual-comb ranging is a time-domain measurement, which means that only a single photodiode is enough to obtain the necessary ranging information. The time-domain signal may be processed using various methods, which include digital processing with Hilbert transform \cite{coddington_rapid_2009}, or alternatively, it can be processed directly in the time-domain with cost-effective time-to-digital converter (TDCs) chips developed for LiDAR applications, or similar approaches \cite{wright_two-photon_2021}. In contrast to WLI, dual-comb ranging offers motion-free and very rapid optical delay sweeps between the reference and probe arms, equivalent to more than 10 km/s. Such high-speed delay scanning is compatible with the rapid motion of the beam scanners. This particularly becomes true when 1-GHz or higher pulse repetition rate dual-comb lasers are employed, as they are directly compatible with optical delay sweeps in 10-100 kHz, while still satisfying the aliasing condition necessary for interferogram sampling.

In this paper, we explore 3D profiling using a prototyped 1-GHz single-cavity dual-comb laser. To leverage the coherence of the two pulse trains, we perform heterodyne beating between the scattered light from the sample and the reference to obtain coherent signal amplification, which then allows for high measurement dynamic range. The signals are digitized and processed using digital signal processing methods. By leveraging the pulsed nature of the dual-comb laser, efficient for non-linear processes, such as supercontinuum generation, further enhancements in resolution and precision could be achieved \cite{camenzind_broadband_2025}.

\section{Experimental Setup}

\begin{figure}[t] 
\centering
\includegraphics[width=\textwidth]{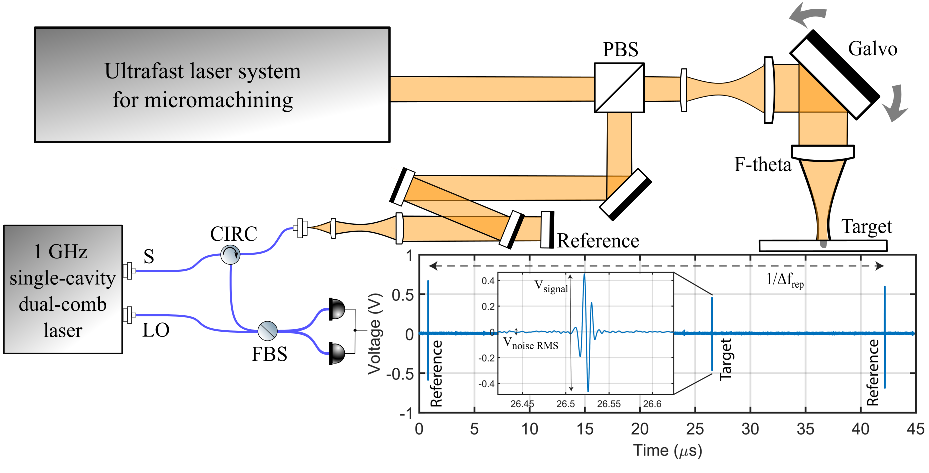} 
\caption{Experimental setup. S – signal, LO – local oscillator, CIRC – circulator, FBS – fiber beam splitter, PBS -polarizing beam splitter. Inset shows example dual-comb LiDAR signal.}
\label{fig:exp_setup}
\end{figure}

Fig.~\ref{fig:exp_setup} shows our setup and a typical signal obtained for a single scanner position. We integrated a recently reported single-cavity dual-comb laser \cite{pupeikis_ultra-low_2024} operating at 1056 nm center wavelength, with 18 nm FWHM bandwidth and $f_{rep}=1$ GHz laser repetition rate into a laser micromachining station. The same dual-comb laser was also employed for long-range dual-comb ranging \cite{camenzind_long-range_2025}. For laser micromachining, we used a Duetto 1064 nm picosecond laser from Time-Bandwidth Products. Because of wavelength overlap, the two lasers were combined via polarization.

The coaxially combined laser beams are then sent into a beam scanning system (intelliSCAN14, SCANLAB), after which the beams are focused with a 100 mm focal length f-theta lens on the sample. In this configuration, the lateral resolution is limited by the aperture of the beam scanner. We estimate the focused mode size to be around 20 µm, corresponding to a full width at half maximum (FWHM) of approximately 12 µm, which sets the practical limit for the lateral resolution of the measurement. The dual-comb laser light backscattered from the sample surface is collected and mixed with a local oscillator to coherently amplify the returning light. The measurement setup was estimated to be sensitive to sub-nW average power of returning light, while up to 60 mW have been sent towards the target indicating ranging sensitivity up to 77 dB. The interferograms were digitized with a 250 megasamples per second (MS/s) sampling card and acquisitions were triggered on the scanner motion. In this proof-of-concept demonstration, the data was processed offline, but the data acquisition card used is compatible with real-time signal processing as we have previously reported in the context of dual-comb range tracking application with the dual-comb laser used here \cite{camenzind_long-range_2025}.

\section{Experimental Results}
\begin{figure*}[t] 
\centering
\includegraphics[width=\textwidth]{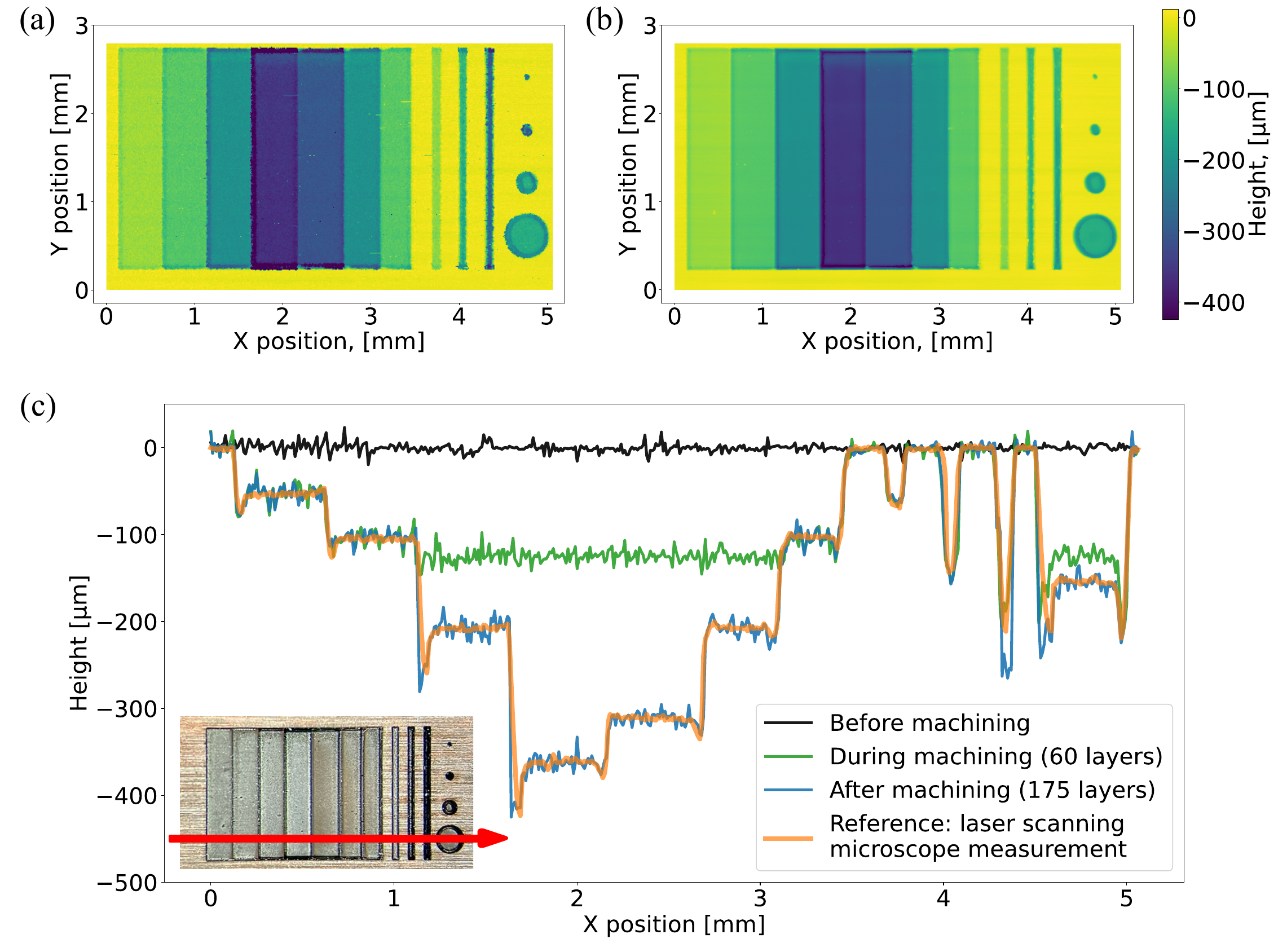} 
\caption{ A test structure was machined in 175 layers. A 3D profile of the surface was measured at three different layers: before the machining started, an intermediate measurement after machining 60 layers and the final measurement after machining all 175 layers. 3D profile after 175 layers: (a) measured in-situ with dual-comb ranging, (b) measured with external laser scanning confocal microscope. (c) Example cross-section of the 3D profiles along the red arrow, including prior, during and after the machining process.}
\label{fig:Profile_2}
\end{figure*}

The sample was profiled in-situ using dual-comb LiDAR at three different stages: prior to machining, during machining (after 60 layers of machining), and after machining (after 175 layers of machining). While the conditions for all three measurements were the same, the machining process was stopped after the 175 axial steps, achieving the desired geometry. The scan utilized a 550 x 300 point grid with a pixel spacing of 10 µm at $\Delta f_{rep}=6.6$ kHz repetition rate difference between the two combs, which determined the effective update rate and corresponds to a single measurement acquisition time of 152 µs. The measurement speed was constrained by the bandwidth of the sampling card. A higher $\Delta f_{rep}$ results in a broader spectrum of the dual-comb LiDAR signal with frequencies above the 125 MHz limit that can be digitized without aliasing at the 250 MS/s sampling rate.
The entire profile scan, including surface scanning and the acquisition of five clean measurements at each lateral position, took approximately three minutes.

Fig.~\ref{fig:Profile_2}a illustrates the depth profile obtained at the conclusion of the machining process. For comparison, the same part was subsequently profiled using an external Laser Scanning Microscope (LSM) (Fig.~\ref{fig:Profile_2}b). We used LSM from Keyence, VK-X130K with a 10-x magnification objective, providing a total 240-x magnification.

A representative cross section of the sample profile before and during machining is shown in Fig.~\ref{fig:Profile_2}c. The depth profile obtained using the dual-comb LiDAR setup aligns closely with the LSM results, with a slight advantage in measuring high-aspect ratio holes. Dual-comb ranging can be considered part of a broader framework of Coherence Scanning Interferometry (CSI) techniques that use interferometric ranging for surface profiling. CSI techniques are known for their compatibility with high-aspect ratio structures~\cite{ma_topography_2024}.

Moreover, the surface smoothness of the measurement results differs noticeably between the two methods, with the dual-comb LiDAR scan exhibiting stronger fluctuations. These fluctuations are partially attributed to speckle-induced errors, which will be discussed in a later section. However, the apparent difference in surface smoothness between the two methods is further amplified by the way the LSM measurement was processed. The LSM measurement in Fig.~\ref{fig:Profile_2}b was performed over multiple sections separately at a high transverse resolution. By the default settings of the LSM software the sections were consequently stitched together, and an averaging in the transverse direction was performed to get a scan at the final output resolution. This artificially smoothens the surface and makes the contrast between the two scans appear more extreme.

\begin{figure}[h] 
\centering
\includegraphics[width=\textwidth]{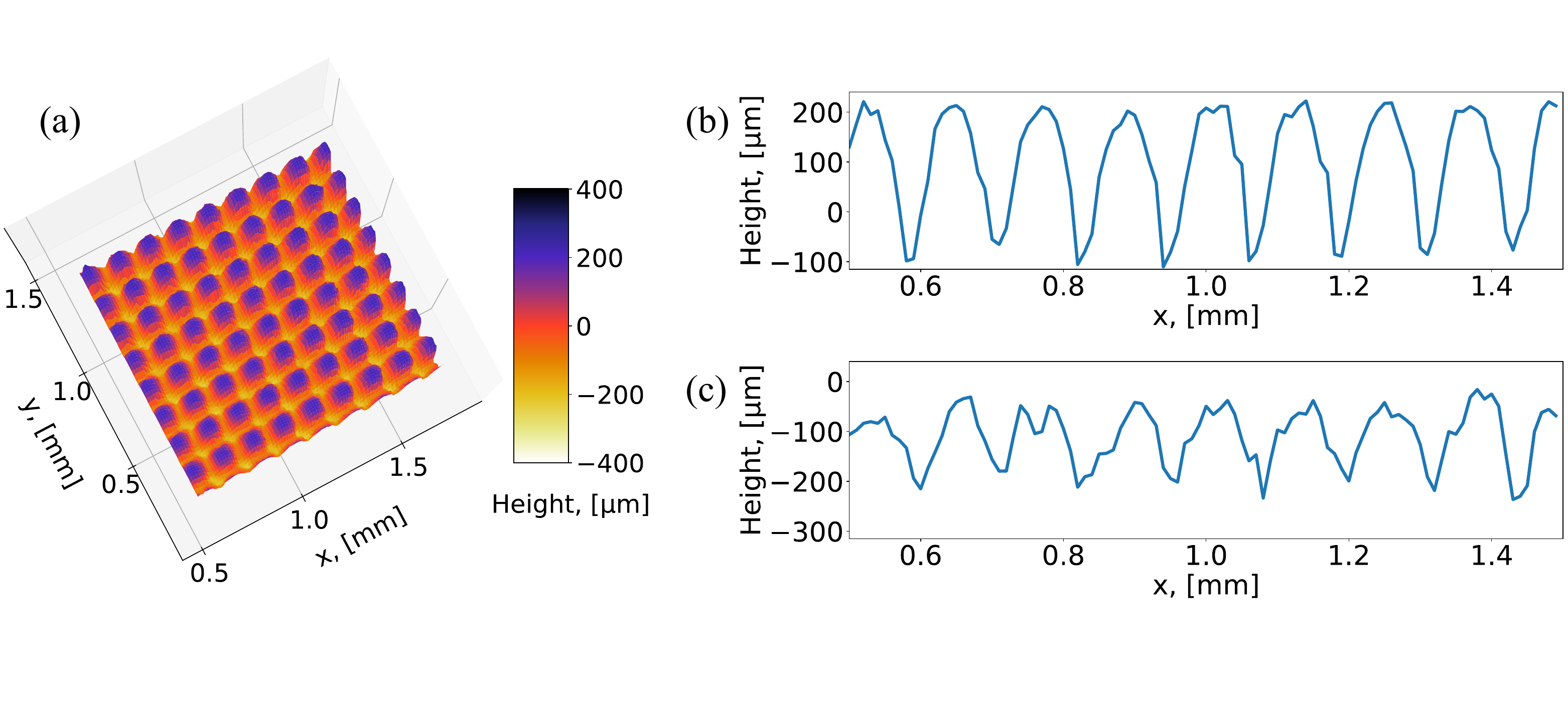} 
\caption{a) LiDAR measurement of a machined double-sine wave profile, b) Cross-section along the peaks, c) Cross-section along the valleys.}
\label{fig:double-sine}
\end{figure}

Another sample with a pre-machined double sine wave profile featuring steep surface angles was scanned to further evaluate the system’s capabilities. The scan, represented in Fig.~\ref{fig:double-sine}, was conducted with a resolution of 100 points per millimeter. Fig.~\ref{fig:double-sine}a) shows an overview of this profile, while b) and c) present cross-sections in the y-direction along the highest and lowest points, respectively.

Developing an optimal laser micromachining process for this microstructure required iterative refinement, alternating between the laser machine and an inspection tool like the LSM. Such iterative optimization processes are common in industrial micromachining but can be labor-intensive. In-situ diagnostics offer a means to periodically measure the ablation depth during the fabrication process while preserving the position and referencing of the workpiece, thereby enabling much more efficient and effective optimization of the process. Our dual-comb ranging setup is compatible with such in-situ measurements.

Before the main experiment described above, as a first evaluation of the dual-comb LiDAR precision, we collected 100 measurements per lateral position on a non-cooperative stationary target. Here we collected the data using a 5 GS/s oscilloscope and could set a higher $\Delta f_{rep}=20$ kHz resulting in a 50 µs acquisition time per measurement. The standard deviation of the distance measurements was used as an uncertainty metric, yielding an average uncertainty of 680 nm across all lateral positions, without averaging multiple measurements per position. As expected in the absence of drifts, the depth measurement precision scales with the square root of the number of averages per lateral position. In case of 5 averages it reached a 310 nm average standard deviation and for 10 averages 220 nm. Based on this, we decided to take five measurements per point for the samples in Fig.~\ref{fig:Profile_2} and Fig.~\ref{fig:double-sine} to obtain a higher precision. As discussed in the next section, the single point measurement uncertainty in these profile measurements is indeed in sub-µm range, however the final precision is limited by the speckle effect caused by a rough surface.

In another experiment using the same dual-comb laser, a single measurement time-of-flight precision of 100 nm was demonstrated at a long range with a cooperative (highly reflective) target at a 5-kHz update rate \cite{camenzind_long-range_2025}.

\section{Discussion}
One apparent feature of the LiDAR measurements in Fig.~\ref{fig:Profile_2}c is the relatively strong fluctuation of the measured profile compared to the expected precision. The theoretically achievable precision of the system depends on the signal-to-noise ratio (SNR) and is limited by the Cramer-Rao Lower Bound, with a single measurement standard deviation given by \cite{kay_fundamentals_1993}
\begin{equation}
    \delta = \frac{c}{2\cdot B\cdot\sqrt{2\cdot SNR}}
\end{equation}

The theoretical limit for the axial resolution is given by $\frac{c}{2\cdot B}$.

A FWHM optical bandwidth of $B=4.84$ THz results in an estimated lower bound for a single point measurement uncertainty of $\delta=\frac{21.9 \textbf{µm}}{\sqrt{SNR}}$ and a resolution of approximately 31 µm. Here, the SNR is defined for time-domain signal power:

\begin{equation}
    SNR = \frac{V_{pp}^2}{V_{noise}^2}
    \label{eq:SNR}
\end{equation}
where $V_{pp}$ it the peak-to-peak voltage and $V_{noise}$ is the rms noise voltage.

With a mean SNR of 1000 observed for the machined surface measurement, the lower bound for single point measurement uncertainty is 0.69 µm. For each lateral position on the surface, five measurements were collected and their standard deviation was calculated. The standard deviations of these single point measurements, averaged over each machined section (1 to 7) in Fig.~\ref{fig:profile 2 sections}, range from 0.7 to 0.85 µm, closely approaching the estimated lower bound.

Based on the demonstrated sub-micrometer single point measurement precision  of the system, we conclude that the transverse fluctuations in the measured profile are a consequence of surface roughness and not due to statistical uncertainty in single point measurements. The measurement is influenced by the surface roughness in two ways: (1) the surface is not smooth, so true fluctuations of the surface profile are expected, and (2) there is a contribution from the speckle effect introducing an additional error caused by the distortion of the signal reflected from a rough surface.

\subsection{Estimation of Surface Roughness and Speckle Induced Uncertainty}

In any coherent laser-ranging method, high spatial frequency surface roughness can impact precision. A finite spot size of the beam at the measurement point means that a rough surface cannot always be laterally resolved. With coherent illumination of the sample, the photodiode samples a portion of the speckle pattern originating from interference between multiple reflections within the illuminated region. This interference introduces additional uncertainty in the measured depth. The effect on axial precision is comparable across all coherent ranging methods such as OCT, WLI, FMCW LiDAR, and dual-comb LiDAR. Confocal approaches, although not relying on temporal coherence, still experience speckle-related fluctuations, but generally to a smaller extent.

The relation between surface roughness and measurement precision has been studied extensively in previous works \cite{dainty_laser_1975,pavlicek_theoretical_2003,pavlicek_white-light_2008, baumann_speckle_2014}. The analytical result using first order statistics is:
\begin{equation}
    \delta z=\frac{1}{\sqrt{2}}\sqrt{\frac{1}{r}}\sigma_h
    \label{eq:delta}
\end{equation}
where $\sigma_h$ is the rms surface roughness, and

\begin{equation}
    r=\frac{I}{\langle I \rangle}
    \label{eq:r}
\end{equation}

is the ratio of the individual speckle intensity $I$ to the mean intensity of the speckle field $\langle I \rangle$ of a rough planar surface.

\begin{figure}[h]
\begin{center}
\includegraphics[width=0.75\linewidth]{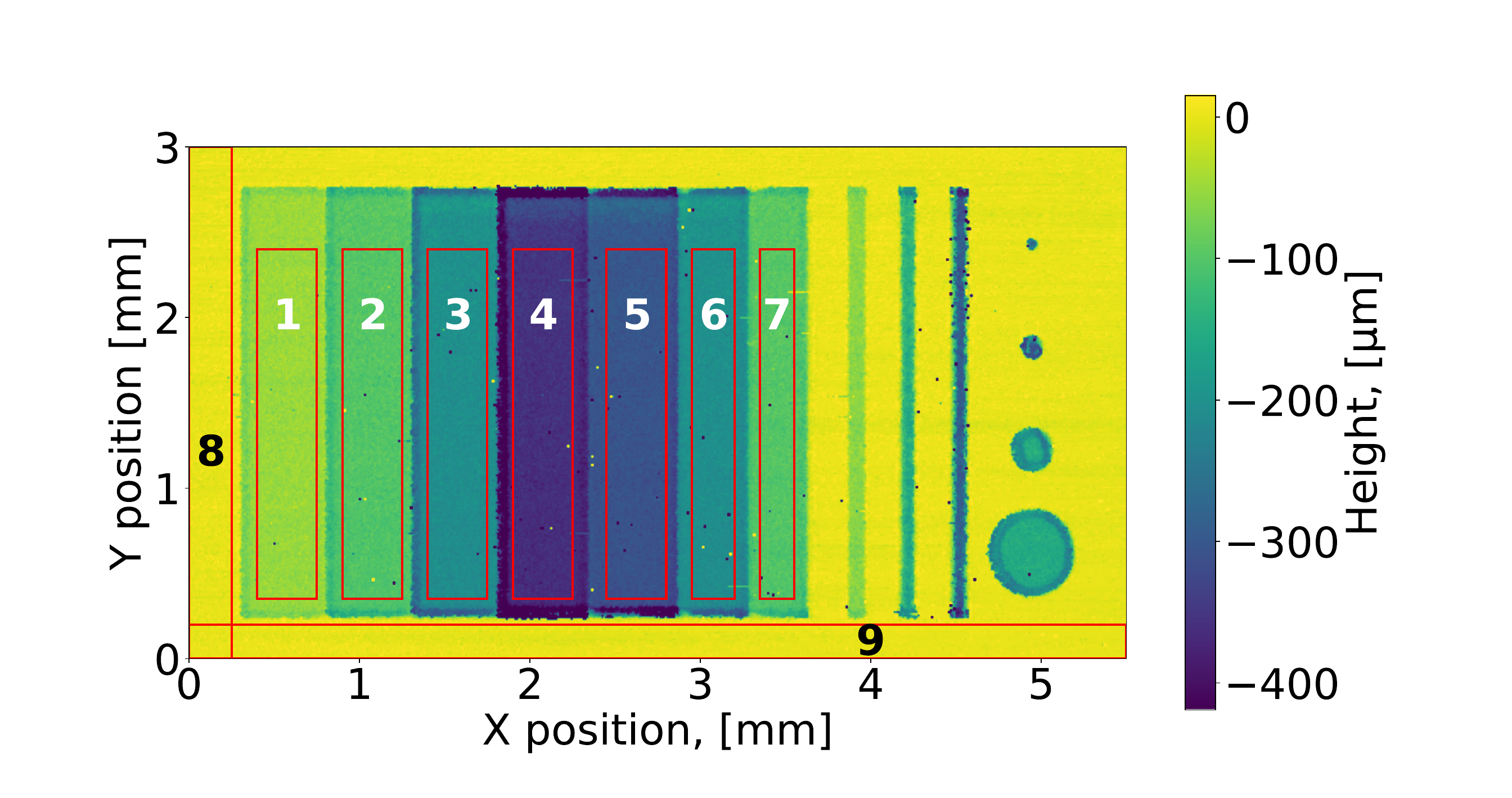}
\caption{ Nine sections of the machined profile where the surface roughness was analyzed.}
\label{fig:profile 2 sections}
\end{center}
\end{figure}

We estimate the distribution of $r$ for the speckle field using $V^2_{pp}$ of the measured photodiode signal. Using this in combination with the eq.~\ref{eq:delta} we can calculate the ratio:
\begin{equation}
    \alpha=\langle \delta z \rangle/\sigma_h
    \label{eq:alpha}
\end{equation}
 between the expectation value of the speckle-induced uncertainty $\langle \delta z\rangle $ and $\sigma_h$ (see the appendix). This has been done for different machined (sections 1-7) and non-machined (sections 8 and 9) planar sections shown in Fig.~\ref{fig:profile 2 sections}. Because there is no statistical dependence between the true position and the measurement error of a single point we can approximate the variance of the height distribution of a measured planar profile as:
\begin{equation}
    \sigma_z^2 = \sigma_h^2+\langle \delta z \rangle^2
    \label{eq:variance}
\end{equation}

For each of the planar sections labeled 1-9 in Fig.~\ref{fig:profile 2 sections} we calculated the variance $\sigma_z$ of the measured height and estimated the ratio $\alpha$ (see the appendix). Using eq.~\ref{eq:alpha} and \ref{eq:variance} we calculate the expectation value of the measurement uncertainty $\langle \delta z \rangle$ and the rms surface roughness $\sigma_h$. The results are shown in Table~\ref{tab:speckle}.

\begin{figure}[ht]
\begin{center}
\includegraphics[width=0.8\linewidth]{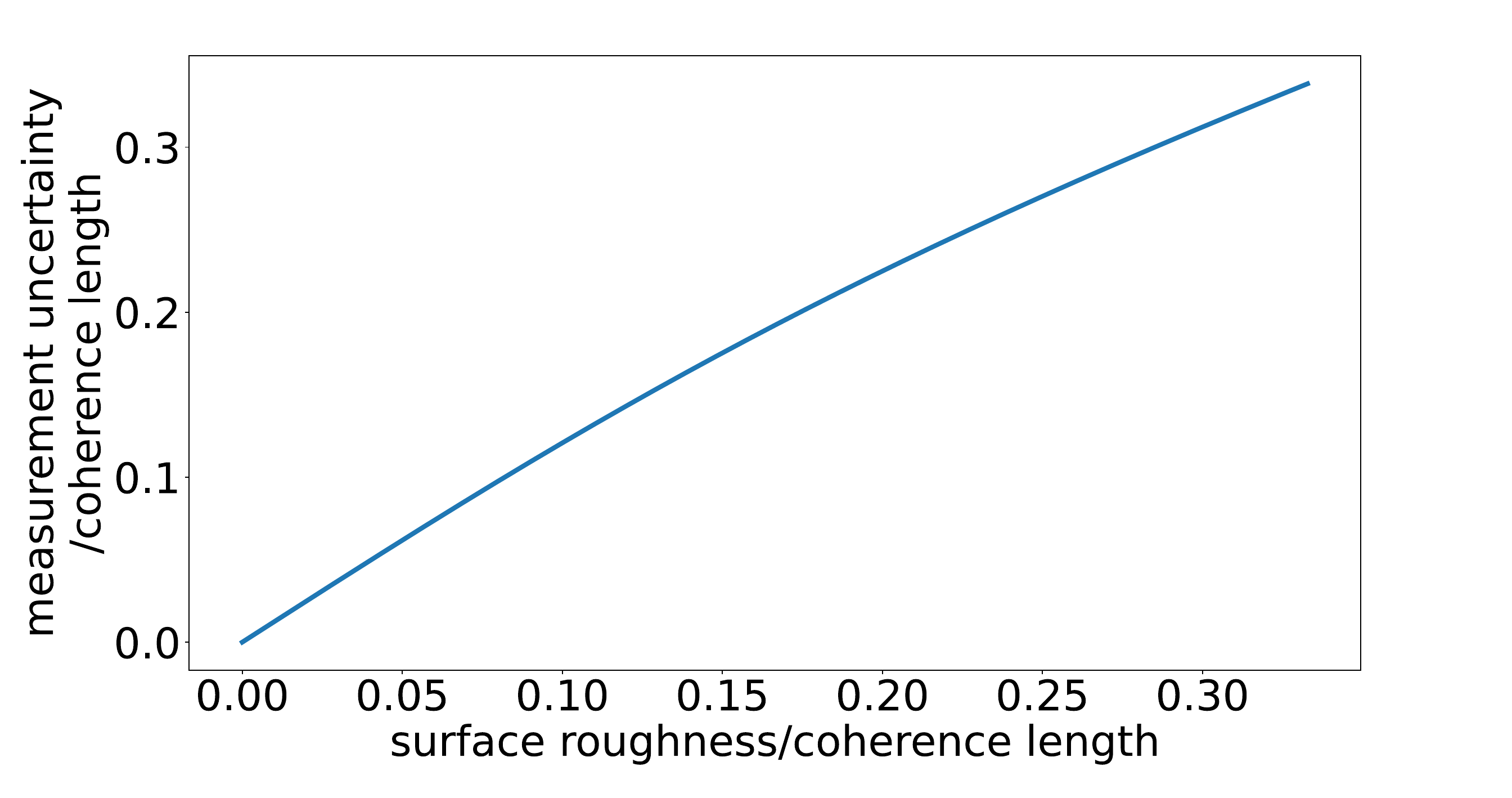}
\caption{ Numerically calculated dependence of speckle induced measurement uncertainty from the surface roughness in units of the coherence length of the laser source.}
\label{fig:numerical error}
\end{center}
\end{figure}

It is also possible to numerically estimate (see the appendix) the relation of speckle induced measurement uncertainty and the surface roughness shown in Fig.~\ref{fig:numerical error} for a more general case of coherent ranging using polychromatic light sources~\cite{dainty_laser_1975,pavlicek_theoretical_2003}. We see that they are approximately on the same scale, which can also be observed in Table~\ref{tab:speckle}.

\begin{table}[ht]
\centering
\begin{tabular}{c|ccccccccc}
\multicolumn{1}{c}{} & \multicolumn{9}{c}{\textbf{Section}} \\ 
\hline
 & 1 & 2 & 3 & 4 & 5 & 6 & 7 & 8 & 9 \\ 
\hline
\textbf{\textit{$\alpha$}} & 1.06 & 1.07 & 1.08 & 1.05 & 1.05 & 1.06 & 1.06 & 1.41 & 1.46 \\ 
$\sigma_z$, [\textmu m] & 7.6 & 7.6 & 7.4 & 7.6 & 7.2 & 7.2 & 7.1 & 5.4 & 4.1 \\ 
$\sigma_h$, [\textmu m] & 5.2 & 5.2 & 5.0 & 5.2 & 4.9 & 4.9 & 4.9 & 3.1 & 2.3 \\ 
$\langle \delta z \rangle$, [\textmu m] & 5.5 & 5.6 & 5.4 & 5.5 & 5.2 & 5.2 & 5.2 & 4.4 & 3.4 \\ 
\end{tabular}
\caption{rms surface roughness $\sigma_h$ and expected speckle induced error $\langle\delta z \rangle$ estimations for sections 1-9 in Fig.~\ref{fig:profile 2 sections}}
\label{tab:speckle}

\end{table}

The relation in Fig.~\ref{fig:numerical error} is derived for a Gaussian spectrum and the corresponding $1/e$ coherence length. In case of soliton pulses, a similar relation is expected, possibly with rescaled coherence length units because of a different spectral shape.


\begin{figure}[ht] 
\centering
\includegraphics[width=\textwidth]{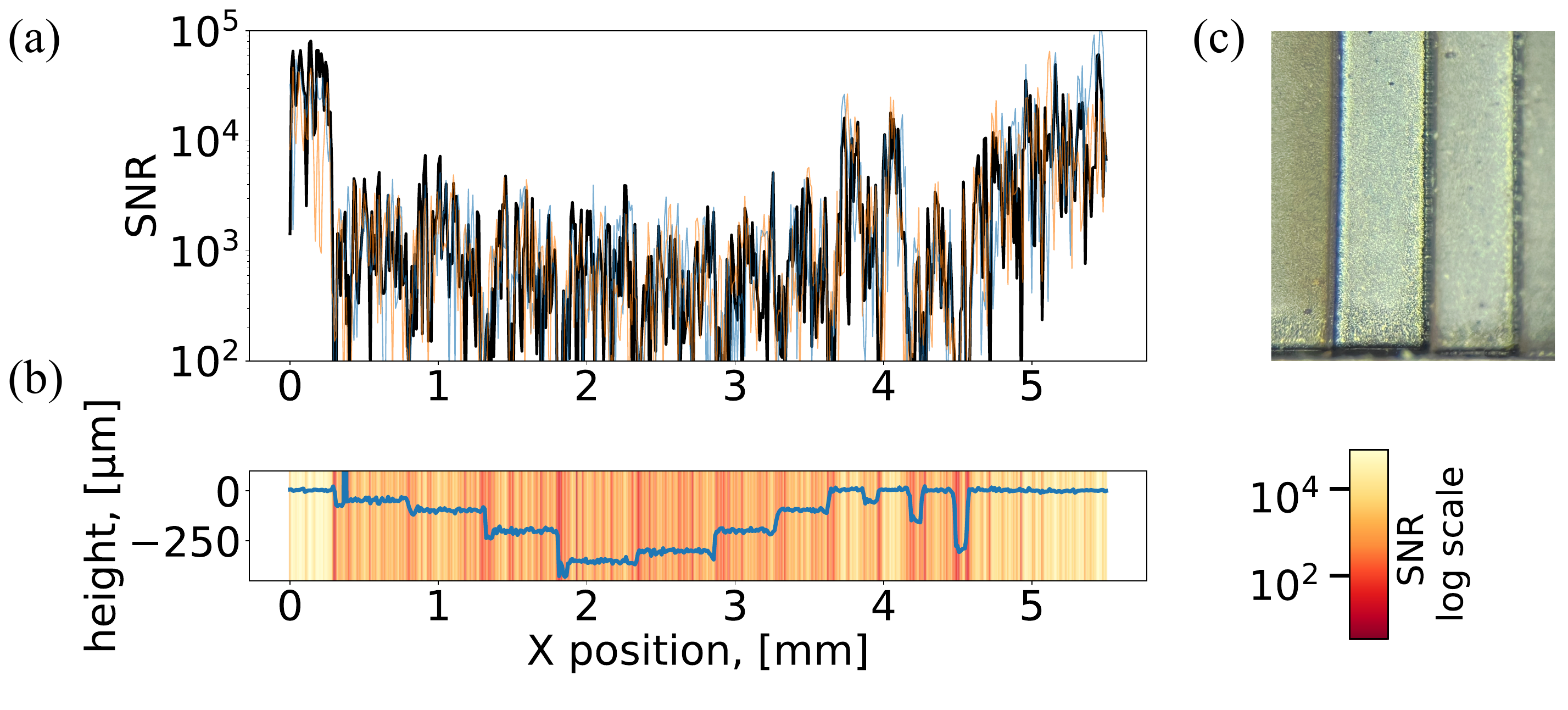} 
\caption{(a) SNR profiles extracted along three different cross-sections of the machined surface. The different colored lines correspond to individual cross-sections, demonstrating that the SNR pattern is consistent across the machined surface. A bold black line is shown as a representative example. (b) Measured height profile and SNR color map along the cross-section, matched with positions of (a), (c) Example of the machined surface with visible graining caused by a drift of the focus in case of an imperfect machining process.}
\label{fig:SNR}
\end{figure}

\subsection{System performance under reduced SNR conditions}

The increased surface roughness from the machining process also reduces the reflectivity, thus influencing the measurement not only through the speckle effect but also through a reduced SNR value. Other factors can also reduce reflectivity, such as the redeposition of removed metal particles on the surface, oxidation of the surface, or other types of surface contamination. Still, even with reduced SNR, for most rough surfaces, the speckle effect remains the dominant source of error.

As shown in Fig.~\ref{fig:SNR}(a), the SNR of machined sections is significantly lower than that of non-machined sections. A mean SNR of 1000 was recorded for the machined sections, corresponding to a 10x–100x reduction in reflectivity. Despite this decrease, the system continues to function effectively with no significant limitations.

\section{Conclusions}
In conclusion, we have demonstrated the integration of coaxial dual-comb LiDAR into a laser micromachining station, enabling 3D profiling during the micromachining process. This integration facilitates in-situ non-destructive testing (NDT) and evaluation of machined parts, reducing the effort required for both process development and control. At $\Delta f_{rep}=6.6$ kHz, a scan of 165000 lateral positions with five measurements at each was completed in three minutes, which can be further reduced by taking fewer measurements per point, increasing the acquisition rate, and optimizing the scanning of the surface. At a mean SNR of 1000 for machined surfaces, the standard deviation over multiple measurements of a single lateral position was 0.8 µm. However, the precision is limited by the speckle induced error, which is on the same scale as the estimated surface roughness. This additional error was estimated to be 5.4 µm on average for the machined surfaces with an estimated rms surface roughness of 5.0 µm. This precision, limited by surface roughness, is therefore insufficient for applications that require profilometry with an accuracy finer than the surface texture itself. However, for the system’s primary role in process development and qualification, where verifying the overall geometry is more relevant, it provides the necessary feedback with sufficient precision and allows to estimate the surface quality. Also the system can provide valuable feedback for focus position correction during machining.

Compared to confocal LSM, the dual-comb LiDAR offers slightly better capability in measuring high-aspect-ratio features. Additionally, unlike OCT techniques, it eliminates the need for reference arm delay matching, making it particularly suitable for profiling large parts ($>> 1$ cm). Beyond serving as a replacement for traditional 3D profilers, this technique can provide auxiliary functionalities such as autorefocusing of the machining laser on the sample and has the potential to enable active machining control. Practical implementation challenges include addressing optical isolation to protect the photodetector from high-power machining laser light and developing methods to combine different wavelengths within a single system to enable integration into a wider range of laser machining setups. These considerations will be crucial for the advancement of this technology toward broader industrial adoption.

Future developments could explore supercontinuum generation to further enhance the performance of the LiDAR system \cite{camenzind_broadband_2025}. By using nonlinear photonic crystal fibers, the comb spectrum can be broadened to 100 THz optical bandwidth, which is expected to improve the axial resolution, precision, and simultaneously reduce the sensitivity to speckle. Such bandwidth expansion could potentially improve resolution by more than an order of magnitude, making the system particularly suitable for applications such as depth control for in-glass machining. In addition, a broader spectrum enables hyperspectral LiDAR capabilities, allowing for material classification and layer differentiation in, for example, semiconductor inspection. However, the associated increase in optical bandwidth proportionally increases the RF bandwidth of the interferograms, demanding faster data acquisition to avoid aliasing at high sampling rates.

\appendix
\section*{Appendix A.\hspace{1em}Estimation of Surface Roughness and Speckle Induced Uncertainty}
\label{app1}

\begin{figure}[ht]
\begin{center}
\includegraphics[width=0.75\linewidth]{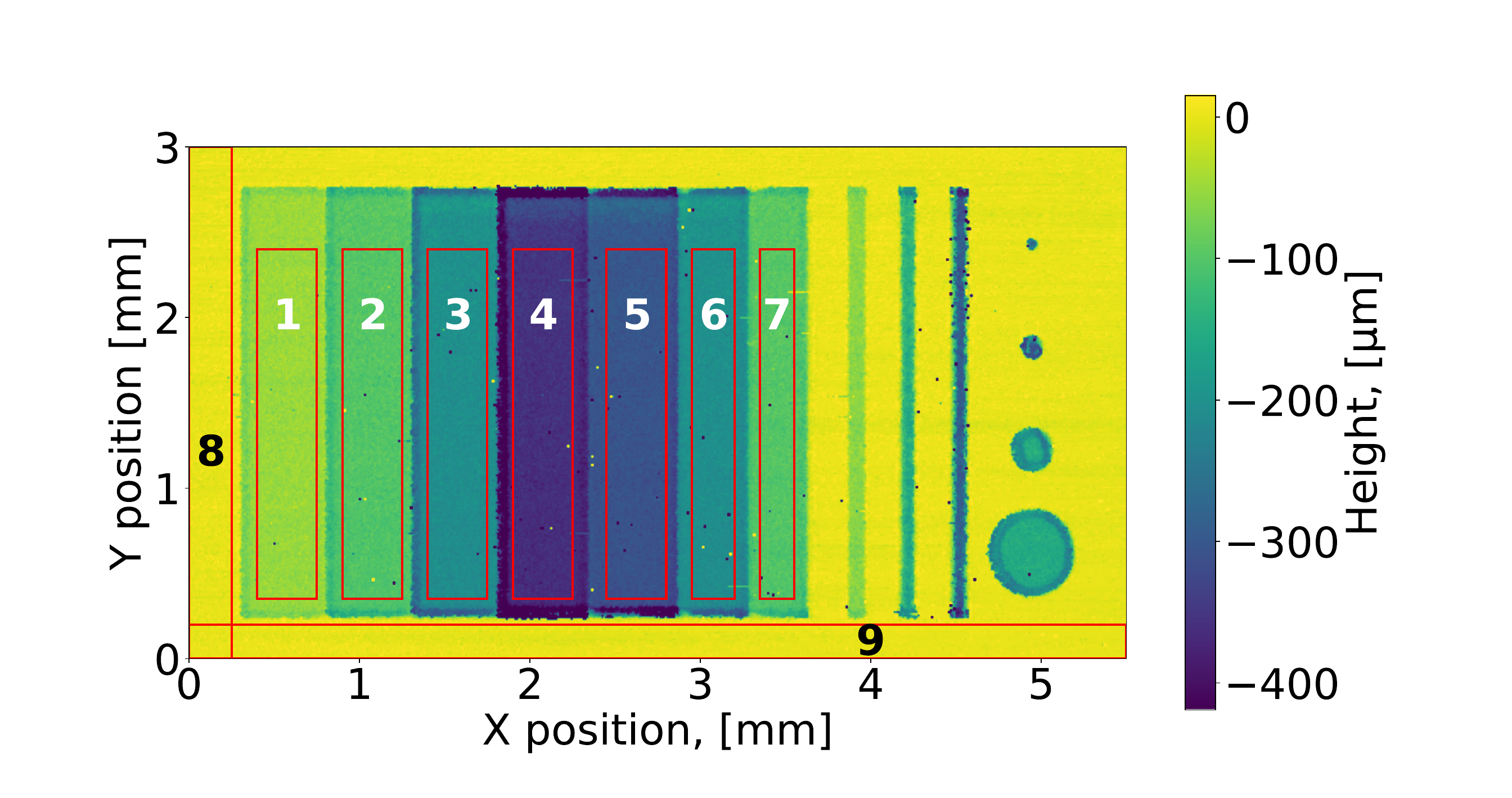}
\caption{ Nine sections of the machined profile where the surface roughness was analyzed.  }
\label{fig:Apx_profile 2 sections}
\end{center}
\end{figure}

In any coherent laser-ranging method, high spatial frequency surface roughness can impact precision. A finite spot size of the beam at the measurement point means that a rough surface cannot always be laterally resolved. With coherent illumination of the sample, the photodiode samples a portion of the speckle pattern originating from interference between multiple reflections within the illuminated region. This interference introduces additional uncertainty in the measured depth.

This contribution of the speckle effect to the measurement uncertainty of coherent ranging methods has been studied in multiple works
\cite{dainty_laser_1975,pavlicek_theoretical_2003,pavlicek_white-light_2008, baumann_speckle_2014}, based on analytical derivations and numerical simulations. Many approaches use first-order statistics to estimate the uncertainty of a range measurement for one single speckle:


\begin{equation}
    \delta z=\frac{1}{\sqrt{2}}\sqrt{\frac{\langle I \rangle}{I}}\sigma_h
    \label{eq:Apx_delta_z}
\end{equation}

Here $\sigma_h$ is the rms surface roughness, $I$ is the intensity of the captured light from a single speckle and $\langle I \rangle$ the average over the whole speckle field of a planar rough surface. The speckle induced uncertainty $\delta z$ of a single measurement scales linearly with the surface roughness $\sigma_h$ and is inversely proportional to $\sqrt{r}$, with:
\begin{equation}
    r=\frac{I}{\langle I \rangle}
    \label{eq:Apx_r}
\end{equation}
being the ratio of speckle intensity to the mean intensity of the speckle field.

\begin{figure}[htbp]
\centering
\begin{subfigure}{0.49\textwidth}
    \centering
    \includegraphics[width=1.0\linewidth]{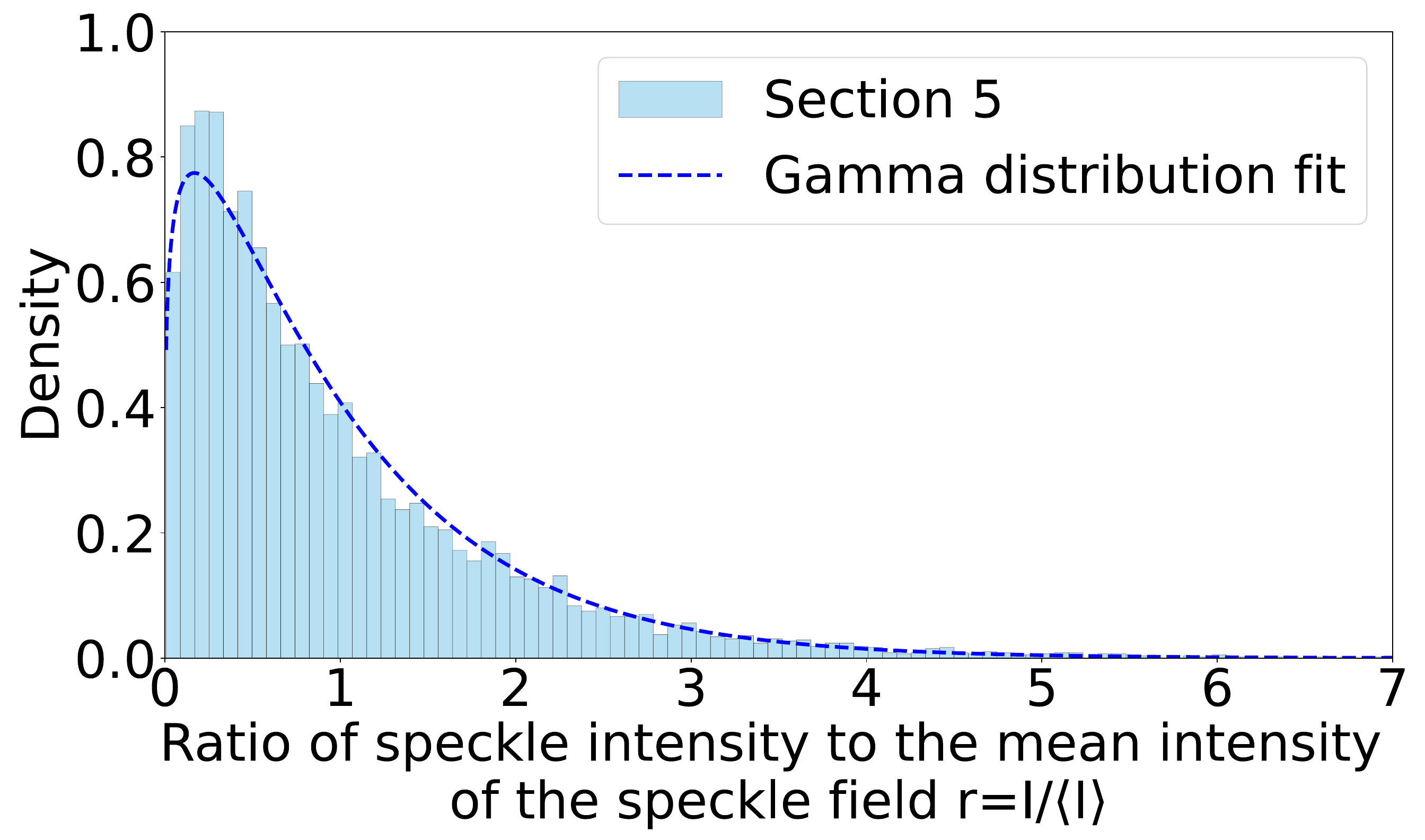}
    \caption{Section 5 with the fit parameter $M=1.2$}
    \label{fig:Apx_intensity5}
\end{subfigure}
\hfill
\begin{subfigure}{0.49\textwidth}
    \centering
    \includegraphics[width=1.0\linewidth]{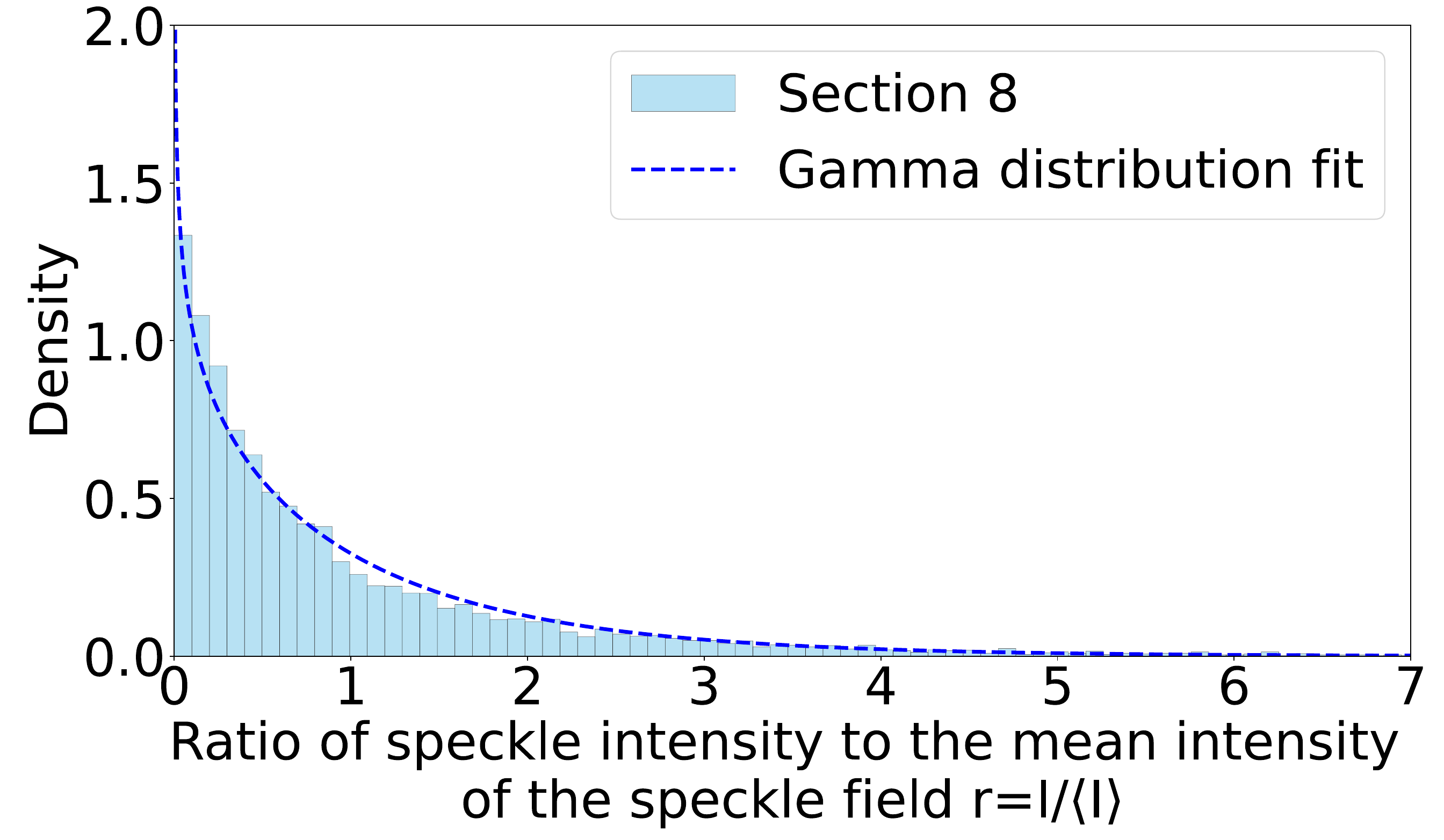}
    \caption{Section 8 with the fit parameter $M=0.81$}
    \label{fig:Apx_intensity8}
\end{subfigure}

\caption{The distribution of the ratio $r$ of individual speckle intensity to the mean intensity in a speckle field.}
\label{fig:Apx_intensity_distribution}
\end{figure}


The distribution of $r$ for polychromatic sources has also been theoretically estimated and validated using simulations \cite{dainty_laser_1975,pavlicek_theoretical_2003,pavlicek_white-light_2008}. Over a planar speckle field, the value of $r$ is shown to follow a gamma distribution \cite{dainty_laser_1975,pavlicek_white-light_2008}: 

\begin{equation}
    p(r) = \frac{M^M r^{M-1}}{\Gamma(M)} \exp(-Mr)
\end{equation}

The statistics of the measured data can be fitted well using this gamma distribution with a fitting parameter $M$. As an estimate for the relative intensity of the reflected light the $V_{pp}^2$ of the photodiode signal was used.

The scan in Fig.~\ref{fig:Apx_profile 2 sections} was used to evaluate the surface roughness and its effects on  precision. Nine planar surfaces were analyzed separately. For each section the mean intensity was calculated, and a gamma distribution was fitted to estimate the probability density function of $r=\frac{I}{\langle I \rangle}$. Two examples of the distribution and the fit are shown in Fig.~\ref{fig:Apx_intensity_distribution} for a machined section 5 and a non-machined section 8.

We call the ratio between the surface roughness $\sigma_h$ and the expectation value $\langle \delta z \rangle$ for the measurement uncertainty $\alpha$. Using the gamma distribution fit and the eq. (\ref{eq:Apx_delta_z}), an expectation value of the measurement uncertainty 

\begin{equation}
    \langle \delta z \rangle = \alpha \cdot \sigma_h
\end{equation}
can be calculated.

After finding the value for $\alpha$ and calculating the standard deviation $\sigma_z$ of the measured surface profile for each of the nine selected planar sections, the rms surface roughness $\sigma_h$ and the expected uncertainty can be estimated using

\begin{equation}
    \sigma_z^2 = \sigma_h^2 + \langle \delta z \rangle^2.
\end{equation}
The results are shown in the table below:

\begin{table}[ht]
\centering
\begin{tabular}{c|ccccccccc}
\multicolumn{1}{c}{} & \multicolumn{9}{c}{\textbf{Section}} \\ 
\hline
 & 1 & 2 & 3 & 4 & 5 & 6 & 7 & 8 & 9 \\ 
\hline
\textbf{M} & 1.2 & 1.17 & 1.16 & 1.22 & 1.2 & 1.19 & 1.15 & 0.81 & 0.78 \\ 
\textbf{\textit{$\alpha$}} & 1.06 & 1.07 & 1.08 & 1.05 & 1.05 & 1.06 & 1.06 & 1.41 & 1.46 \\ 
$\sigma_z$, [\textmu m] & 7.6 & 7.6 & 7.4 & 7.6 & 7.2 & 7.2 & 7.1 & 5.4 & 4.1 \\ 
$\sigma_h$, [\textmu m] & 5.2 & 5.2 & 5.0 & 5.2 & 4.9 & 4.9 & 4.9 & 3.1 & 2.3 \\ 
$\langle \delta z \rangle$, [\textmu m] & 5.5 & 5.6 & 5.4 & 5.5 & 5.2 & 5.2 & 5.2 & 4.4 & 3.4 \\ 
\end{tabular}
\caption{rms surface roughness $\sigma_h$ and expected speckle induced error $\langle\delta z \rangle$ estimations for sections 1-9 in Fig.~\ref{fig:Apx_profile 2 sections}}
\label{tab:Apx_speckle}
\end{table}

\begin{figure}[ht]
\begin{center}
\includegraphics[width=0.8\linewidth]{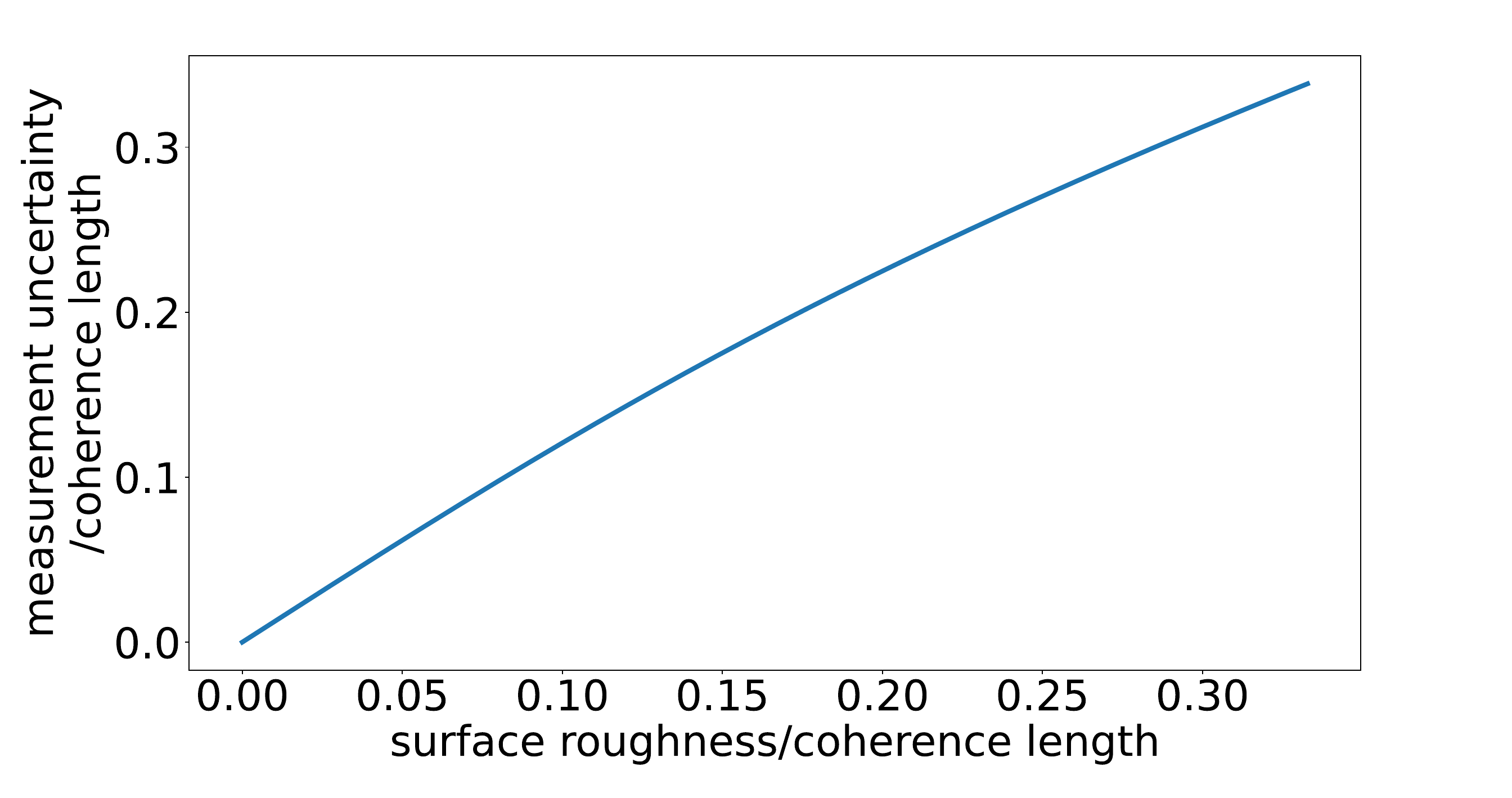}
\caption{ Numerically calculated dependence of speckle induced measurement uncertainty from the surface roughness in units of the coherence length of the laser source.}
\label{fig:Apx_numerical error}
\end{center}
\end{figure}

The relation of speckle induced measurement uncertainty and the surface roughness can also be numerically estimated. The parameter $M$ in the gamma distribution is estimated to be~\cite{dainty_laser_1975,pavlicek_white-light_2008}:

\begin{equation}
M = \sqrt{1 + \left(\frac{\sigma_h}{l_c}\right)^2}
\label{eq:Apx_M}
\end{equation}
in case of a Gaussian spectrum where $l_c$ is the coherence length. For a soliton pulse similar behavior is expected, possibly with some prefactor before $\left(\frac{\sigma_h}{l_c}\right)^2$. Using this, we can numerically calculate the dependence between $\frac{\sigma_h}{l_c}$ and $\frac{\langle \delta z \rangle}{l_c}$, shown in Fig.~\ref{fig:Apx_numerical error}. We see that they are approximately on the same scale, which can also be observed in Table~\ref{tab:Apx_speckle}.

However, the theoretical model discussed above is only valid for the range where the surface roughness covers multiple wavelengths of the light source used and does not exceed 1/4 of the coherence length of the optical bandwidth. We already see some discrepancy for non-machined sections 8 and 9 for which the surface roughness is too close to the wavelength. 
According to eq. (\ref{eq:Apx_M}) the fitting parameter $M$ must be greater than $1$, however, for sections 8 and 9 it is smaller.

\begin{backmatter}
\bmsection{Funding}
This project has received funding from Innovation Booster Photonics under Grant Nr. IB-P 23.022.

\bmsection{Acknowledgment}
The authors thank S. L. Camenzind for useful discussions. The authors acknowledge experimental support from Matthias Julius. 

\bmsection{Disclosures}
LL (E), BW (E), and JP (E) declare a partial employment with K2 Photonics.

\bmsection{Data Availability}
Data underlying the results presented in this article are available in the ETH Zurich Research Collection Library.

\end{backmatter}

\bibliography{Bibliography}

\end{document}